# Relativistic Correction to the First Moment of the Spin-Dependent Structure Function of the Deuteron $\Gamma_1^D(Q^2)$ in the Light-Cone Formalism

**F.F. Pavlov**

*St. Petersburg State Polytechnical University*

*Russia, 195251, St. Petersburg, Polytechnicheskaya, 29*

*E-mail: pavlovfedor@mail.ru, f.pavlov@tuexph.stu.neva.ru*

**Abstract**

The deuteron is considered as a superposition of two-nucleon Fock states with the invariant mass depending on the relative momentum in a proton–neutron pair. The condition of the transversality of the polarization vectors should be imposed at the Fock component level and these vectors depend on the invariant mass of the Fock component. Such "running" longitudinal polarization vector was not used in early estimates of relativistic effects. The technique for the calculation of the average helicity of the proton in the deuteron has been considered in the light-cone variables. A receipt has been proposed for the consistent calculation of relativistic nuclear corrections to the average helicity of the proton in the deuteron and to the first moment of the spin-dependent structure function of the deuteron.

# I. Introduction

The convenient nonrelativistic consideration of the deuteron is inapplicable to effects associated with high momenta in the deuteron and cannot provide the description of the entire set of experimental data. For this reason, relativistic effects owing to the high-momentum component in the deuteron are of current interest and require adequate theoretical description.

In this work, the wave function of the deuteron is approximated by a proton-neutron Fock state using the previously developed methods of the relativistic field theory in the light-cone variables that are successfully applied in quantum chromodynamics for description of spin phenomena at the exclusive production of vector mesons in the deep-inelastic scattering of leptons on protons [1].

The deuteron is considered as a superposition of two-nucleon Fock states with the invariant mass depending on the relative momentum in a proton–neutron pair. The condition of the transversality of the polarization vectors should be imposed at the Fock component level and these vectors depend on the invariant mass of the Fock component. Such "running" longitudinal polarization vector was not used in early estimates of relativistic effects.

The technique for calculating nuclear corrections to the spin structure function of the deuteron is topical, because there is no consistent procedure for taking into account relativistic corrections to the spin structure function of the deuteron, which is extracted from the data for the proton and deuteron.

## II. Amplitude of single electron–deuteron scattering

The triangular Feynman diagram of the single electron–deuteron scattering (Fig. 1) includes the vertex of the interaction of a photon with one of the nucleons in the deuteron (where the second nucleon is a spectator on the mass shell), the vertex function of the decay of the deuteron into the proton and neutron in the initial state, and the vertex function of the deuteron in the final state. For Feynman diagrams, the deuteron as a particle with the spin $S = 1$ is a massive vector meson (*Proca particle*, *massive photon*, *etc*.). However, in contrast to convenient diagrams in quantum electrodynamics or in the theory of weak interaction, where a fermion is created and an antifermion is absorbed at electroweak vertices (or an antifermion is created and a fermion is absorbed), the absorption of the deuteron at the deuteron–proton–neutron (*Dpn*) vertex is accompanied by the creation of two fermions – the proton and neutron. However, the proton *p* and antineutron $\bar{n} \equiv a_n$ can be considered as basis particles and the deuteron can be considered as the bound state $D = p\bar{a}_n$, so that the absorption of the deuteron at the vertex $Dpn = D\bar{p}a_n$ is accompanied by the creation of the fermion *p* and absorption of the fermion $a_n$. In this case, the vertex is written in the conventional form $\bar{\psi}_p \Gamma_\mu \psi_{a_n} D^\mu$, where $D^\mu$ is the operator of the deuteron field described by a 4-vector and $\Gamma_\mu$ is the vertex function. As is known, the current $\bar{\psi}\gamma_\mu\psi$, where $\gamma_\mu$ are the Dirac matrices, is transformed as a 4-vector under Lorentz transformations. Since the parity of $a_n$ is opposite to the parity of the neutron, the parity of $\bar{\psi}_p\gamma_\mu\psi_{a_n}$ is positive, as should be for the deuteron. In terms of the propagation of the proton and antineutron $a_n$, the diagram in Fig. 1 has the form of a usual fermion loop.

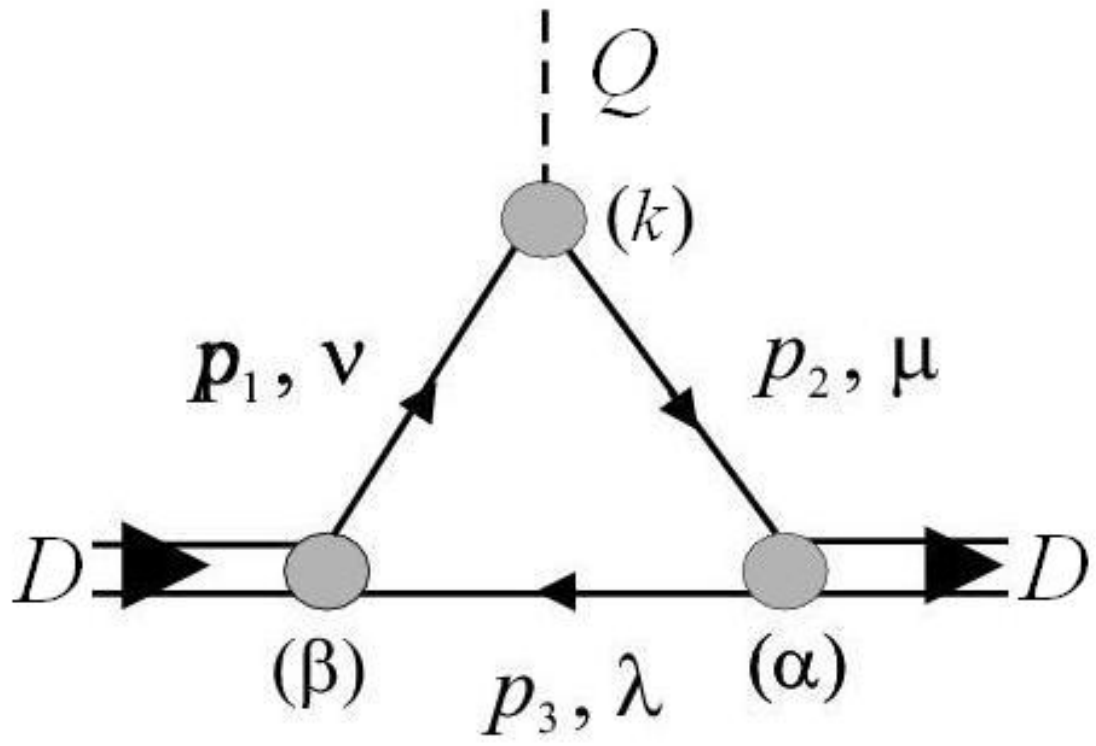

**Fig. 1.** Feynman diagram for the deuteron.

With the standard Feynman rules, the vertex part of the amplitude of the process corresponding to Fig. 1 can be represented in the form

$$A_k = (-1)\int \frac{d^4 p_3}{(2\pi)^4} \frac{Sp\left\{ i\left(\Gamma_\beta V_\beta^{(\rho)}\right) i\left(-\hat{p}_3 + m\right) i\left(\Gamma_\alpha^* V_\alpha^{(\rho)*}\right) i(\hat{p}_2 + m) i O_k i(\hat{p}_1 + m)\right\}}{\left(p_3^2 - m^2 + i\varepsilon\right)\left(p_2^2 - m^2 + i\varepsilon\right)\left(p_1^2 - m^2 + i\varepsilon\right)}. \tag{1}$$

Here, $p_1$ and $p_2$ are the 4-momenta of the protons; integration is performed with respect to the 4-momentum of the neutron $p_3$; the integration contour is closed around the pole of the neutron propagator (masses of all nucleons are *m*); $\hat{p} = p_\mu \gamma_\mu$; repeated indices imply summation; $\Gamma_\beta$ is the total vertex function of the transition of the deuteron into a proton–neutron pair in the initial state; $\Gamma_\alpha^*$ is the total vertex function of the deuteron in the final state; $V_\beta^{(\rho)}$ and $V_\alpha^{(\rho)*}$ are the polarization four-vectors of the deuteron in the helicity representation, respectively; and $\rho, \rho' = \pm 1, 0$ is the helicity of the deuteron. The vertex $O_k$ of the interaction of nucleons with a photon has the form

$$O_k = F_1^N(Q^2)\gamma_k + \frac{F_2^N(Q^2)}{2m} i\sigma_{k\nu} Q_\nu, \tag{2}$$

where $Q$ is the momentum transfer, $F_{1,2}^N(Q^2)$ are the electromagnetic form factors of the nucleon, and $\sigma_{k\nu} = \frac{i}{2}(\gamma_k \gamma_\nu - \gamma_\nu \gamma_k)$.

## III. Vertex Functions of the Deuteron. Radial Wave Function and its Relation to the Vertex Function

The total vertex function of the transition of the deuteron into a proton-neutron pair has the form

$$\Gamma_\beta = \Gamma_\beta^S G_S(M^2) + \Gamma_\beta^D G_D(M^2) . \qquad (3)$$

Here,

$$\Gamma_\beta^S = \gamma_\beta - \frac{(p_1 - p_3)_\beta}{M + 2m} , \qquad (4)$$

and

$$\Gamma_\beta^D = \frac{M^2 - 4m^2}{4}\gamma_\beta + \frac{M + m}{2}(p_1 - p_3)_\beta , \qquad (5)$$

are the vertex functions of the deuteron for the *S* and *D* states, respectively [1–3]; and $G_{S,D}(M^2)$ are the scalar vertex functions for the *S* and *D* states of the deuteron, respectively, which are related to the radial wave functions of the deuteron $\Phi_{S,D}(M^2)$ [1] as

$$\Phi_{S,D}(M^2) = \frac{G_{S,D}(M^2)}{M^2 - M_D^2} , \qquad (6)$$

where $M$ is the invariant mass of the proton–neutron pair and $M_D$ =1875,6 MeV/$c^2$ is the mass of the deuteron.

## IV. Two-Particle State in the Light-Cone Variables

The light-cone technique was considered in [1, 5–13]. The two-particle state can be considered in the light-cone variables with the internal 4-momenta $p_1$ and $p_3$. The total 4-momentum of the deuteron is $P = p_1 + p_3$. Since $P_+ = p_{1+} + p_{3+}$, it is convenient to introduce the quantities

$$z = p_{1+} / P_+ ,\; 1 - z = p_{3+} / P_+ , \qquad (7)$$

which are the fractions of the momentum of the system that are carried by particles 1 and 3, respectively.

Then,

$$p_{1-} = \frac{m_1^2 + \mathbf{p}_{1\perp}^2}{2zP_+}, \; p_{3-} = \frac{m_3^2 + \mathbf{p}_{3\perp}^2}{2(1-z)P_+}. \tag{8}$$

The square of the invariant mass of two-nucleon Fock states is

$$M^2 = P^2 = (p_1 + p_3)^2 =$$

$$= \frac{m_1^2 + \mathbf{p}_{1\perp}^2}{z} + \frac{m_3^2 + \mathbf{p}_{3\perp}^2}{1-z} - (\mathbf{p}_{1\perp} + \mathbf{p}_{3\perp})^2. \tag{9}$$

The transverse momentum $\mathbf{P}_\perp = \mathbf{p}_{1\perp} + \mathbf{p}_{3\perp}$ describes the motion of the system as a whole. The relative transverse momentum $\mathbf{k}$ for two initial nucleons is defined through the relations

$$\mathbf{p}_{1\perp} = \mathbf{k} + z\mathbf{P}_\perp, \tag{10}$$

$$\mathbf{p}_{3\perp} = -\mathbf{k} + (1-z)\mathbf{P}_\perp. \tag{11}$$

It follows from Eqs. (9)-(11) at $m_1 = m_3 = m$ that

$$M^2 = \frac{\mathbf{k}^2 + m^2}{z(1-z)}. \tag{12}$$

In terms of the light-cone variables, the 4-momentum of two-nucleon Fock states with the invariant mass $M$ is represented in the form

$$P = (P_+, P_-, \mathbf{P}_\perp) = \left( P_+, \frac{M^2 + \mathbf{P}_\perp^2}{2P_+}, \mathbf{P}_\perp \right). \tag{13}$$

In terms of the light-cone variables, the helicity states for two-nucleon Fock states with the invariant mass $M$ are described by the longitudinal ($\rho = 0$) 4-polarization

$$V^{(\rho=0)} = \frac{1}{M}\left( P_+, \frac{-M^2 + \mathbf{P}_\perp^2}{2P_+}, \mathbf{P}_\perp \right), \tag{14}$$

and by the transverse ($\rho = \pm 1$) 4-polarization in the light-cone variables

$$V^{(\rho=\pm1)} = \left( 0, \frac{\left(\mathbf{P}_\perp \cdot \mathbf{e}^{(\rho=\pm1)}\right)}{P_+}, \mathbf{e}^{(\rho=\pm1)} \right). \tag{15}$$

Here, the transverse circular unit vectors have the convenient form

$$\mathbf{e}^{(\rho=\pm1)} = -\frac{1}{\sqrt{2}}\left(\pm\mathbf{e}_1 + i\mathbf{e}_2\right), \qquad (16)$$

where $\mathbf{e}_1$ and $\mathbf{e}_2$ are the unit vectors along the $x$ and $y$ axes, respectively.

It should be emphasized that $M \neq M_D$ in Eq. (14). The polarization vector of the longitudinal state in the relativistic case inevitably depends on the invariant mass $M$ of the proton-neutron pair. Such a longitudinal polarization vector of the two-nucleon Fock state with the invariant mass has not yet been used in early estimates of relativistic effects.

If the pair moves along the $z$-axis, i.e., $\mathbf{P}_\perp = 0$ or $(P_x, P_y) = (0,0)$, then and the helicity states $V_\beta^{(\rho=\pm1)} = \left(0,\ 0,\ \mathbf{e}^{(\rho=\pm1)}\right)$ can be used as transverse polarization vectors. For the longitudinal state,

$$V_\beta^{(\rho=0)} = \frac{1}{M}\left(P_+, -\frac{M^2}{2P_+}, 0, 0\right). \qquad (17)$$

## V. Matrix Element of the Electromagnetic Vertex of the Deuteron Current

The matrix element of the electromagnetic vertex of the deuteron current $j_k$ is related to $A_k$ as $F_k = \langle D'|j_k|D\rangle = -iA_k$.

As was shown in [8, 11–13], the use of the plus component of the deuteron current $j_+$ ($F_+ = \langle D'|j_+|D\rangle$) in the infinite-momentum system in the special Breit reference frame ($Q_+ = 0$) provides a correct space-time description of relativistic effects (impossibility of pair production from vacuum).

The plus component of amplitude (1) determines the normalization condition for the charge form factor of the deuteron at zero photon momentum transfer: $F_+ = 2P_+F_1^N(0)(w_S + w_D) = 2P_+$, where $w_S$ and $w_D$ are the probabilities of the $S$ and $D$ states in the deuteron, respectively, and $w_S + w_D = 1$.

The calculation of the trace in amplitude (1) was discussed in detail in [1]. Omitting the stages of the calculation of the one-loop integral, the matrix element of the electromagnetic vertex $F_+$ of the deuteron current in the light-cone variables can be reduced to the form

$$F_+ = \frac{1}{2(2\pi)^3}\int \frac{dzd^2\mathbf{k}}{z^2(1-z)} \sum_{\lambda,\nu} \frac{[\bar{\mathrm{v}}(p_3,\lambda)V_\alpha^{(\rho)*}\Gamma_\alpha^* u(p_1,\nu)][\bar{u}(p_1,\nu)\gamma_+ u(p_1,\nu)][\bar{u}(p_1,\nu)V_\beta^{(\rho)}\Gamma_\beta \mathrm{v}(p_3,\lambda)]}{\left(M^2-M_D^2\right)^2} =$$

$$= \frac{1}{2(2\pi)^3}\int \frac{dzd^2\mathbf{k}}{z^2(1-z)} \sum_{\lambda,\nu} \Phi_{\lambda\nu}^* \left[\bar{u}(p_1,\nu)\gamma_+ u(p_1,\nu)\right]\Phi_{\nu\lambda}\,, \tag{18}$$

where

$$\Phi_{\nu\lambda} = \frac{\bar{u}(p_1,\nu)V_\beta^{(\rho)}\Gamma_\beta \mathrm{v}(p_3,\lambda)}{M^2-M_D^2} =$$

$$= \left[\bar{u}\left(p_1,\nu\right)V_\beta^{(\rho)}\Gamma_\beta^S \mathrm{v}\left(p_3,\lambda\right)\right]\Phi_S(M^2) + \left[\bar{u}\left(p_1,\nu\right)V_\beta^{(\rho)}\Gamma_\beta^D \mathrm{v}\left(p_3,\lambda\right)\right]\Phi_D(M^2)\,. \tag{19}$$

Here, $u(p_1,\nu)$ is the spinor of the proton (the incoming fermion in the Feynman diagram) with the momentum $p_1$ and helicity $s=\nu/2$, $\nu=\pm 1$ [1, 7, 9], $\mathrm{v}(p_3,\lambda)$ is the spinor of the neutron (the incoming antifermion in the Feynman diagram) with the momentum $-p_3$ and helicity $-s=\lambda/2$, $\lambda=\pm 1$ [1, 7, 9], $\bar{u}=u^+\gamma_0$, and $\bar{\mathrm{v}}=\mathrm{v}^+\gamma_0$; $\gamma_\pm = \frac{1}{\sqrt{2}}(\gamma_0 \pm \gamma_3)$.

It is worth noting that the spinors in the light-cone formalism differ from the conventional Dirac spinors only in the spin rotation, which is the known Melosh–Wigner transformation [10, 14].

The analytical form of Eq. (19) will be given in Eqs. (42) and (43). Expressions for the nucleon matrix elements can be found in [1, 7, 9], where spinors in the light-cone formalism were used, in particular,

$$\hat{p}_1 + m = \sum_{\nu=\pm 1} u\left(p_1,\nu\right)\bar{u}\left(p_1,\nu\right), \tag{20}$$

$$-\hat{p}_3 + m = -\sum_{\lambda=\pm 1} \mathrm{v}\left(p_3,\lambda\right)\bar{\mathrm{v}}\left(p_3,\lambda\right), \tag{21}$$

$$\bar{u}(p_1,\nu)\gamma_+ u(p_1,\nu) = 2p_{1+} = 2zP_+\,, \tag{22}$$

$$\bar{u}(p_1,\nu)\gamma_+\gamma_5 u(p_1,\nu) = 2\nu p_{1+} = 2\nu zP_+\,. \tag{23}$$

Formula (18) allows the simple quantum-mechanical interpretation: the deuteron in the spin state described by the polarization vector $V_\beta^{(\rho)}$ with the helicity $\rho$ is represented as the proton-neutron system with the helicities $\nu$ and $\lambda$. After scattering, the proton-neutron system is projected on the proton-neutron system in the spin state described by the polarization vector $V_\alpha^{(\rho')}$ with the helicity $\rho'$. Summation is performed over all intermediate helicities $\nu$ and $\lambda$ and this summation replaces the calculation of Feynman traces.

After necessary transformations of matrix elements (19), the total combination of the *S* state of the vertex function for transverse polarizations of the two-nucleon Fock state with the invariant mass *M* for the helicity $\rho = \pm 1$ can be represented in the form

$$\bar{u}(p_1,\nu)V_\beta^{(\rho=\pm1)}\Gamma_\beta^S \mathrm{v}(p_3,\lambda) = -\frac{m(1+\rho\nu)\delta_{\nu\lambda}}{\sqrt{2z(1-z)}} + \frac{\left[-(1-2z)+\rho\nu\right]k(\rho)\delta_{\nu,-\lambda}}{\sqrt{2z(1-z)}} +$$
$$+\frac{2k(\rho)\left[m(1-2z)\delta_{\nu,-\lambda}+k(-\lambda)\delta_{\nu\lambda}\right]}{(M+2m)\sqrt{2z(1-z)}}, \qquad (24)$$

where $k(\rho) = \sqrt{2}\left(\mathbf{k}\cdot\mathbf{e}^{(\rho)}\right) = -\rho k_1 - ik_2$, $k(-\lambda) = \lambda k_1 - ik_2$; $\mathbf{k} = (k_1, k_2)$, $\delta_{\nu\lambda}$ is the Kronecker delta.

For the longitudinal polarization $\rho = 0$,

$$\bar{u}(p_1,\nu)V_\beta^{(\rho=0)}\Gamma_\beta^S \mathrm{v}(p_3,\lambda) = -2M\sqrt{z(1-z)}\delta_{\nu,-\lambda} - \frac{(1-2z)M}{M+2m}\frac{1}{\sqrt{z(1-z)}}\left[m(1-2z)\delta_{\nu,-\lambda}+k(-\lambda)\delta_{\nu\lambda}\right] \qquad (25)$$

Here, $V_\alpha^{(0)}$ should be treated as the running longitudinal polarization vector depending on invariant mass (12) of a pair of nucleons.

According to the above expressions, the transverse deuteron contains an impurity of states of two nucleons with light-cone helicities $\lambda = \nu = -1$. Similarly, the longitudinal deuteron with the helicity $\rho = 0$ contains not only states of two nucleons with $\lambda + \nu = 0$, but also an impurity of states with $\lambda = \nu = \pm 1$. The difference between the sum of the helicities of the nucleons and the helicity of the deuteron is due to the orbital angular momentum of the pair.

Similarly, for the $D$ state,

$$\bar{u}(p_1,\nu)V_\beta^{(\rho=\pm1)}\Gamma_\beta^D \mathrm{v}(p_3,\lambda)=-\frac{M^2-4m^2}{4}\frac{m(1+\rho\nu)\delta_{\nu\lambda}}{\sqrt{2z(1-z)}}+\frac{M^2-4m^2}{4}\frac{k(\rho)\delta_{\nu,-\lambda}}{\sqrt{2z(1-z)}}\left[-(1-2z)+\rho\nu\right]-$$

$$-\frac{(M+m)k(\rho)}{\sqrt{2z(1-z)}}\left[m(1-2z)\delta_{\nu,-\lambda}+k(-\lambda)\delta_{\nu\lambda}\right], \qquad (26)$$

$$\bar{u}(p_1,\nu)V_\beta^{(\rho=0)}\Gamma_\beta^D \mathrm{v}(p_3,\lambda)=-\frac{\left(M^2-4m^2\right)M}{2}\sqrt{z(1-z)}\delta_{\nu,-\lambda}+$$

$$+\frac{(M+m)M(1-2z)}{2\sqrt{z(1-z)}}\left[m(1-2z)\delta_{\nu,-\lambda}+k(-\lambda)\delta_{\nu\lambda}\right]. \qquad (27)$$

It is noteworthy that there is remarkable compensation of rather complex contributions of the squares of vertex functions (24) and (25) for two-nucleon $S$-wave Fock state with the invariant mass $M$:

$$\sum_{\lambda,\nu}\Phi_{\nu\lambda}^{(\rho=\pm1,0)*}\Phi_{\nu\lambda}^{(\rho=\pm1,0)}=2M^2\left|\Phi_S\left(M^2\right)\right|^2. \qquad (28)$$

It can be seen that, first, the squares of vertex functions (24) and (25) are independent of $z$ and $\mathbf{k}^2$, but depend only on the radial variable $M^2$ and, second, they are the same for all helicities of the two-nucleon Fock state, as should be for the purely $S$ state of the deuteron. The latter property is valid only if running longitudinal polarization vector (14) is used. If the external longitudinal polarization vector defined for the fixed mass of the deuteron $M_D$ was used, this property would be violated. The formal reason is that the longitudinally polarized vector state is mixed in this case with the scalar state, which would violate the relation between the angular symmetry between the states of the deuteron with various helicities.

From the normalization condition for form factors, the normalization conditions for the radial wave functions of the Fock states with a certain invariant mass for $S$ and $D$ waves separately were obtained in the form [1]

$$\frac{1}{(2\pi)^3}\int\frac{dzd^2\mathbf{k}}{z(1-z)}M^2\left|\Phi_S(M^2)\right|^2=\frac{1}{(2\pi)^3}\int d^3\mathbf{p}4M\left|\Phi_S(M^2)\right|^2=w_S, \qquad (29)$$

$$\frac{1}{(2\pi)^3}\int\frac{dzd^2\mathbf{k}}{z(1-z)}2M^2\mathbf{p}^4\left|\Phi_D(M^2)\right|^2=\frac{1}{(2\pi)^3}\int d^3\mathbf{p}8M\mathbf{p}^4\left|\Phi_D(M^2)\right|^2=w_D\,, \tag{30}$$

$$\int\frac{dzd^2\mathbf{k}}{z(1-z)}=\int\frac{4}{M}d^3\mathbf{p}\,, \tag{31}$$

where $\mathbf{p}=(\mathbf{k},p_z)$ is the relative three-dimensional momentum in the deuteron introduced in [15], $p_z=-\frac{1}{2}(1-2z)M$ , $\mathbf{p}^2=\frac{1}{4}M^2-m^2$. It is worth noting that $\langle\mathbf{k}^2\rangle=(2/3)\mathbf{p}^2$.

In the nonrelativistic formalism, the usually used normalization is

$$\int_0^\infty dpp^2\left[\left(\Psi_S(p)\right)^2+\left(\Psi_D(p)\right)^2\right]=w_S+w_D=1\,. \tag{32}$$

The relation between the radial wave functions $\Phi_{S,D}$ and nonrelativistic wave functions $\Psi_{S,D}$ has the form

$$\left|\Phi_S\right|^2=\frac{\pi^2}{2M}\left|\Psi_S\right|^2, \tag{33}$$

$$\left|\Phi_D\right|^2=\frac{\pi^2}{4M\mathbf{p}^4}\left|\Psi_D\right|^2. \tag{34}$$

A number of modern realistic wave functions, e.g., Bonn [16] and Paris [17] wave functions can be used as nonrelativistic wave functions $\Psi_{S,D}$.

We represent the expression (18) as [18, 19]

$$F_+=2P_+\int n^{(1,1)}(z)\,dz=\frac{2P_+}{2(2\pi)^3}\int\frac{dzd^2\mathbf{k}}{z(1-z)}\sum_{\lambda,\nu}\Phi_{\nu\lambda}^{(\rho=1)*}\Phi_{\nu\lambda}^{(\rho=1)}=2P_+\,. \tag{35}$$

Then

$$n^{(1,1)}(z)=\frac{1}{2(2\pi)^3}\int\frac{d^2\mathbf{k}}{z(1-z)}\sum_{\lambda,\nu}\Phi_{\nu\lambda}^{(\rho=+1)*}\Phi_{\nu\lambda}^{(\rho=+1)}=$$

$$=\frac{1}{2(2\pi)^3}\int\frac{d^2\mathbf{k}}{z(1-z)}\Big\{2M^2\left|\Phi_S\left(M^2\right)\right|^2+2M^2\mathbf{p}^2\left[\mathbf{p}^2+(3/2)\mathbf{k}^2\right]\left|\Phi_D\left(M^2\right)\right|^2+$$

$$+4M^2\left[\mathbf{p}^2-(3/2)\mathbf{k}^2\right]\Phi_S\left(M^2\right)\Phi_D\left(M^2\right)\Big\}. \tag{36}$$

It is worth noting that $\langle\mathbf{k}^2\rangle=(2/3)\mathbf{p}^2$.

For the longitudinal polarization $\rho = 0$,

$$n^{(0,0)}\left(z\right) = \frac{1}{2\left(2\pi\right)^3}\int \frac{d^2\mathbf{k}}{z(1-z)}\sum_{\lambda,\nu}\Phi_{\nu\lambda}^{(\rho=0)*}\Phi_{\nu\lambda}^{(\rho=0)} =$$

$$= \frac{1}{2(2\pi)^3}\int \frac{d^2\mathbf{k}}{z(1-z)}\Big\{2M^2\left|\Phi_S\left(M^2\right)\right|^2 + 2M^2\mathbf{p}^2\left[4\mathbf{p}^2 - 3\mathbf{k}^2\right]\left|\Phi_D\left(M^2\right)\right|^2 +$$

$$+ 4M^2\left[3\mathbf{k}^2 - 2\mathbf{p}^2\right]\Phi_S\left(M^2\right)\Phi_D\left(M^2\right)\Big\}. \qquad (37)$$

Knowing the explicit form of $n^{(1,1)}\left(z\right)$ (36) can be found deuteron unpolarized structure functions $F_1^D(x_D)$ and $F_2^D(x_D)$. Within the limit of Bjorken scaling we obtain:

$$F_1^D\left(x_D\right) = \int_{x_D}^{1} \frac{dz}{z} n^{(1,1)}\left(z\right) F_1^N\left(\frac{x_D}{z}\right), \qquad (38)$$

$$F_2^D\left(x_D\right) = \int_{x_D}^{1} dz n^{(1,1)}\left(z\right) F_2^N\left(\frac{x_D}{z}\right). \qquad (39)$$

The eqs. (38)-(39) have a simple interpretation: the probability of finding a parton in the deuteron, carrying a fraction of the deuteron momentum $x_D$, is equal to the product the probability of finding a parton in the nucleon with a fraction of the nucleon momentum $x_N = x_D / z$ and the probability of finding a nucleon with a fraction of the deuteron momentum *z*.

## VI. Relativistic Nuclear Correction to the Average Helicity of the Proton

The estimate of various relativistic corrections is of current interest for the consistent relativistic description of the deuteron. The relativistic expression for the average helicity of the proton in the deuteron is estimated in this work. As is known, the doubled helicity $\langle \nu_p \rangle$ in the nonrelativistic approximation is given by he expression

$$\left\langle \nu_p \right\rangle_{nonrel} = \left\langle S_z \right\rangle = w_S - \frac{1}{2} w_D = 1 - \frac{3}{2} w_D . \tag{40}$$

This expression is significantly different in the relativistic case [18–19]. The average helicity is given by the expression

$$\left\langle \nu_p \right\rangle = \frac{1}{2P_+} \int \frac{d^4 p_3}{(2\pi)^4 i} \frac{Sp\left\{ \left( \Gamma_\beta V_\beta^{(\rho)} \right) \left( -\hat{p}_3 + m \right) \left( \Gamma_\alpha^* V_\alpha^{(\rho)*} \right) (\hat{p}_2 + m) \gamma_+ \gamma_5 (\hat{p}_1 + m) \right\}}{\left( p_3^2 - m^2 + i\varepsilon \right) \left( p_2^2 - m^2 + i\varepsilon \right) \left( p_1^2 - m^2 + i\varepsilon \right)} =$$

$$= \frac{\int \frac{dzd^2\mathbf{k}}{z(1-z)} \sum_{\lambda,\nu} \nu \, \Phi_{\nu\lambda}^{(\rho=1)*} \Phi_{\nu\lambda}^{(\rho=1)}}{\int \frac{dzd^2\mathbf{k}}{z(1-z)} \sum_{\lambda,\nu} \Phi_{\nu\lambda}^{(\rho=1)*} \Phi_{\nu\lambda}^{(\rho=1)}} , \tag{41}$$

where [18]

$$\Phi_{\nu\lambda}^{(\rho=\pm1)} = \frac{1}{(M+2m)\sqrt{2z(1-z)}} \left\{ \left[ -m\left(M+2m\right)(1+\rho\nu) + 2k\left(\rho\right)k\left(-\lambda\right) \right] \delta_{\nu\lambda} + \right.$$

$$\left. + \left[ \left(2z - 1 + \rho\nu\right)\left(M+2m\right) + 2m\left(1-2z\right) \right] k\left(\rho\right) \delta_{\nu-\lambda} \right\} \Phi_S(M^2) +$$

$$+ \frac{1}{4\sqrt{2z(1-z)}} \left\{ \left[ -\left(M^2 - 4m^2\right) m(1+\rho\nu) - 4(M+m) k\left(\rho\right) k\left(-\lambda\right) \right] \delta_{\nu\lambda} + \right.$$

$$\left. + \left[ \left(2z - 1 + \rho\nu\right)\left(M^2 - 4m^2\right) - 4(M+m)m\left(1-2z\right) \right] k\left(\rho\right) \delta_{\nu-\lambda} \right\} \Phi_D(M^2) , \tag{42}$$

$$\Phi_{\nu\lambda}^{(\rho=0)} = \frac{1}{(M+2m)\sqrt{z(1-z)}} \left\{ \left[ -\left(1-2z\right) Mk\left(-\lambda\right) \right] \delta_{\nu\lambda} + \right.$$

$$\left. + \left[ -2z\left(1-z\right) M\left(M+2m\right) - \left(1-2z\right)^2 Mm \right] \delta_{\nu-\lambda} \right\} \Phi_S(M^2) +$$

$$+ \frac{1}{2\sqrt{z(1-z)}} \left\{ \left[ \left(1-2z\right) M(M+m) k\left(-\lambda\right) \right] \delta_{\nu\lambda} + \right.$$

$$\left. + \left[ -z\left(1-z\right) M\left(M^2 - 4m^2\right) + \left(1-2z\right)^2 Mm(M+m) \right] \delta_{\nu-\lambda} \right\} \Phi_D(M^2) , \tag{43}$$

$$\int \frac{dzd^2\mathbf{k}}{z(1-z)} \sum_{\lambda,\nu} \Phi_{\nu\lambda}^{(\rho)*} \Phi_{\nu\lambda}^{(\rho)} = 2(2\pi)^3 . \tag{44}$$

The final expression for the average relativistic helicity of the proton in the deuteron has the form

$$\left\langle \nu_p \right\rangle = \frac{1}{2(2\pi)^3}\int \frac{dzd^2\mathbf{k}}{z(1-z)}\left\{ \frac{2}{(M+2m)z(1-z)}\left\{ \mathbf{k}^2\left[-\left(1-2z\right)M+2m\right]+m^2\left(M+2m\right)\right\}\Phi_S^2(M^2)+\right.$$

$$+\frac{M^2-4m^2}{8z(1-z)}\left\{\mathbf{k}^2\left[-\left(M+2m\right)\left[(1-2z)M+2m\right]-4mM(1-z)\right]+m^2\left(M^2-4m^2\right)\right\}\Phi_D^2(M^2)+$$

$$\left.+\frac{1}{z(1-z)}\left\{\mathbf{k}^2\left[-\left(M+2m\right)\left[(1-2z)M+2m\right]+2mM(1-z)\right]+m^2\left(M^2-4m^2\right)\right\}\Phi_S(M^2)\Phi_D(M^2)\right\},$$

(45)

or

$$\left\langle \nu_p \right\rangle = \frac{1}{(2\pi)^3}\int d^3\mathbf{p}\left\{ \frac{4M}{\left(M+2m\right)\left(\mathbf{k}^2+m^2\right)}\left[2\mathbf{k}^2(p_z+m)+m^2(M+2m)\right]\Phi_S^2(M^2)+\right.$$

$$+\frac{2\mathbf{p}^2M}{\left(\mathbf{k}^2+m^2\right)}\left\{\mathbf{k}^2\left[M(p_z-2m)+4mp_z\right]+2m^2p_z^2\right\}\Phi_D^2(M^2)+$$

$$\left.+\frac{2M}{\left(\mathbf{k}^2+m^2\right)}\left\{\mathbf{k}^2\left[M(2p_z-m)+2mp_z\right]+4m^2p_z^2\right\}\Phi_S(M^2)\Phi_D(M^2)\right\}, \tag{46}$$

where $\mathbf{p}=\left(\mathbf{k},p_z\right)$, $p_z=-\frac{1}{2}\left(1-2z\right)M$ , $\mathbf{p}^2=\frac{1}{4}M^2-m^2$.

Expression (46) can be decomposed into the convenient nonrelativistic part given by Eq. (40) and the relativistic correction $\Delta_{rel}$:

$$\left\langle \nu_p \right\rangle = w_S-\frac{1}{2}w_D+\Delta_{rel}=1-\frac{3}{2}w_D+\Delta_{rel}. \tag{47}$$

Then,

$$\Delta_{rel}=\frac{1}{(2\pi)^3}\int d^3\mathbf{p}\left\{\frac{4M\mathbf{k}^2(2p_z-m)}{\left(M+2m\right)\left(\mathbf{k}^2+m^2\right)}\Phi_S^2(M^2)+\right.$$

$$+\frac{\mathbf{p}^2M}{\left(\mathbf{k}^2+m^2\right)}\left[2\mathbf{k}^2p_z\left(M+4m\right)+8m^2p_z^2+\mathbf{k}^2M\left(M-4m\right)\right]\Phi_D^2(M^2)+$$

$$\left.+\frac{2M}{\left(\mathbf{k}^2+m^2\right)}\left\{\mathbf{k}^2\left[M\left(2p_z-m\right)+2mp_z\right]+4m^2p_z^2\right\}\Phi_S(M^2)\Phi_D(M^2)\right\}. \tag{48}$$

The expression for the average helicity for the relativistic case includes interference contributions from *S* and *D* states. For $\mathbf{k}^2 << m^2$, Eq. (46) transforms to nonrelativistic formula (40) and $\Delta_{rel}=0$. The Bonn and Paris deuteron wave functions are used for the numerical calculation. The results of the calculation of the average helicity are presented in the table. As can be seen in the table, the relativistic correction is less than 1% of the total value. The interference contributions to the relativistic correction from the *S* and *D* states are small, but nonzero in the relativistic case. The contribution of the relativistic correction to the average helicity of the proton in the deuteron is small.

**Average helicities calculated for two approximations.**

| Formula | Wave-function | Part of the wave state | | | Total value |
|---|---|---|---|---|---|
| | | *S* | *D* | *SD* (interference) | |
| **Nonrelativistic case** | | | | | |
| $\langle v_p \rangle_{nonrel} = w_S - \frac{1}{2} w_D$ | Bonn | 1,0 | -0,5 | 0 | 0,93625 |
| | Paris | | | | 0,91345 |
| **Relativistic case (calculation by Eq. (45))** | | | | | |
| $\langle v_p \rangle$ | Bonn | 0,99507 | -0,48690 | $1{,}1252 \cdot 10^{-4}$ | 0,93223 |
| | Paris | 0,99485 | -0,48337 | $4{,}0319 \cdot 10^{-5}$ | 0,90957 |

Note: In the nonrelativistic case, $w_D$ = 0.0425 and 0.0577 for the Bonn and Paris wave functions, respectively.

If average helicity (45) is represented in the form

$$\langle v_p \rangle = \int_0^1 v(z)\,dz\,, \tag{49}$$

the dependence of the integrand $v(z)$ (distribution of the average helicity of the deuteron) on *z* can be estimated [18]. Figures 2–5 shows $v(z)$ obtained with the Bonn wave function of the deuteron [16]. It can be easily seen that peculiar asymmetry appears in the relativistic case in contrast to the nonrelativistic case. In the relativistic case, the nucleon carrying a larger fraction *z* of the momentum of the system makes a larger contribution to the distribution of the average helicity of the deuteron.

According to Fig. 2, the average helicity local in $z$ obtained for the deuteron in the light-cone formalism strongly differs from the prediction of the nonrelativistic formalism.

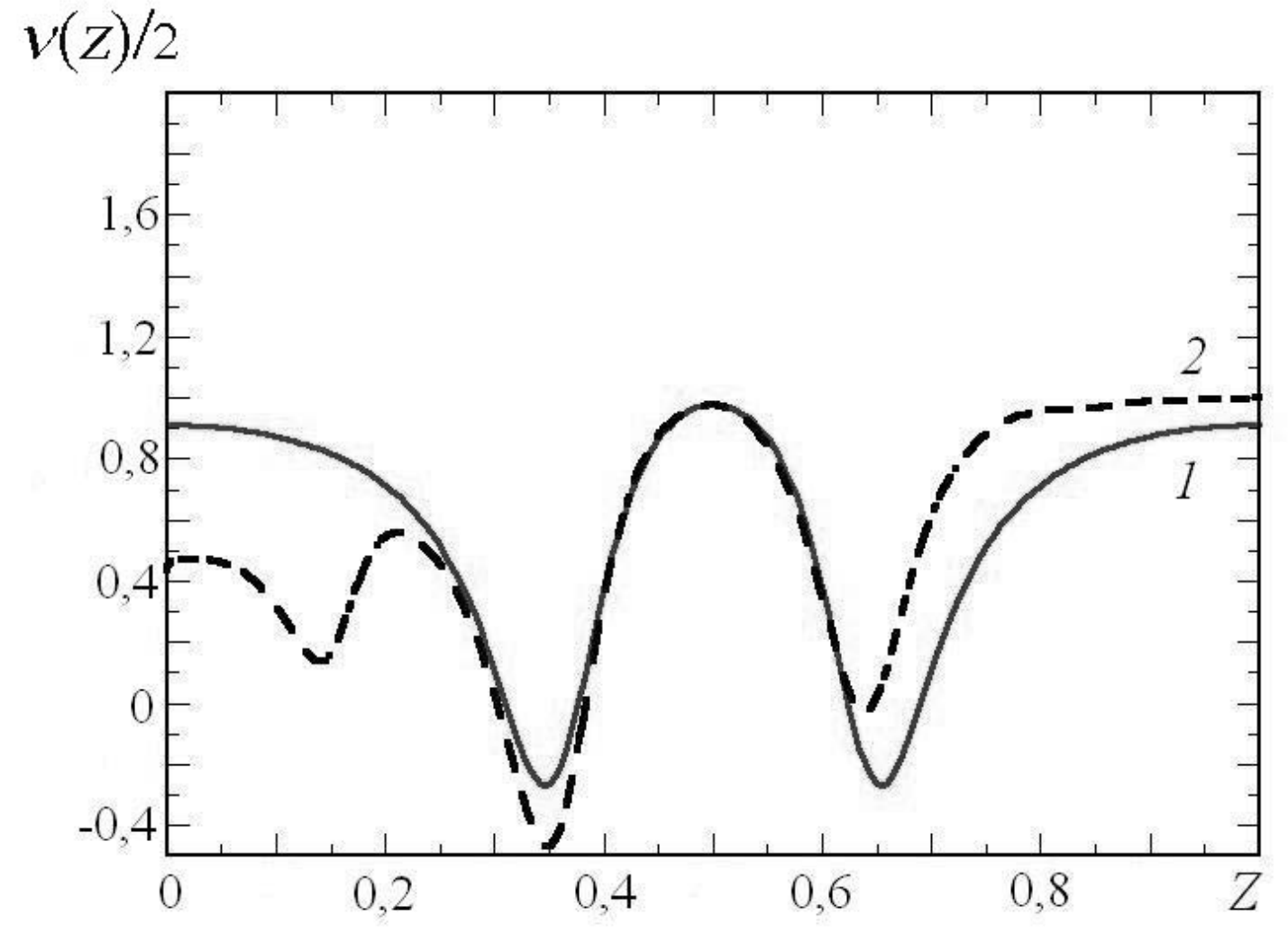

**Fig. 2.** Average helicity $\nu(z)/2$ versus the fraction $z$ of the momentum of the system for the (*1* – solid line) nonrelativistic and (*2* – dashed line) relativistic cases.

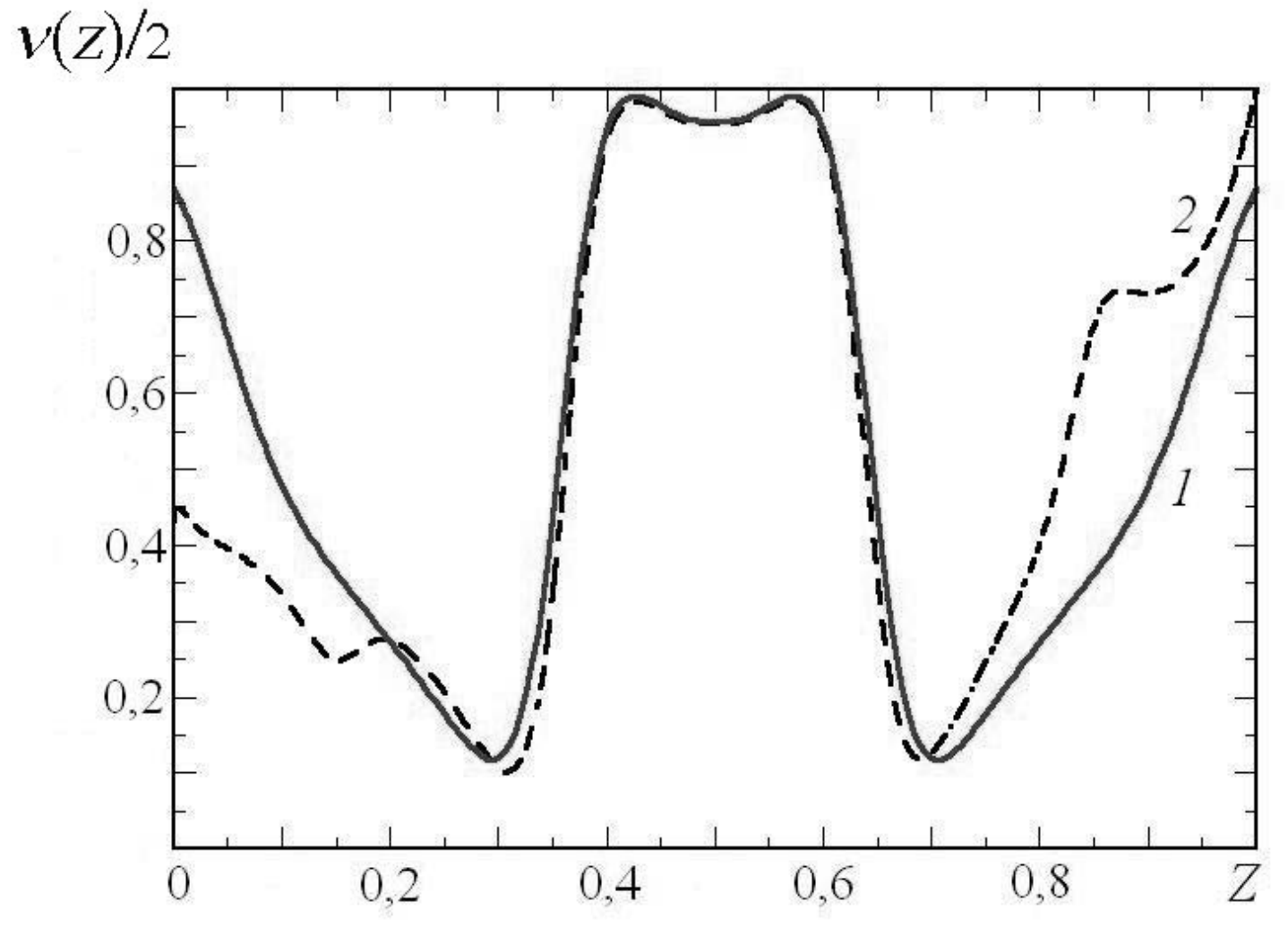

**Fig. 3.** Helicity for the purely *S* state of the deuteron versus the fraction $z$ of the momentum of the system for the (*1* – solid line) nonrelativistic and (*2* – dashed line) relativistic cases.

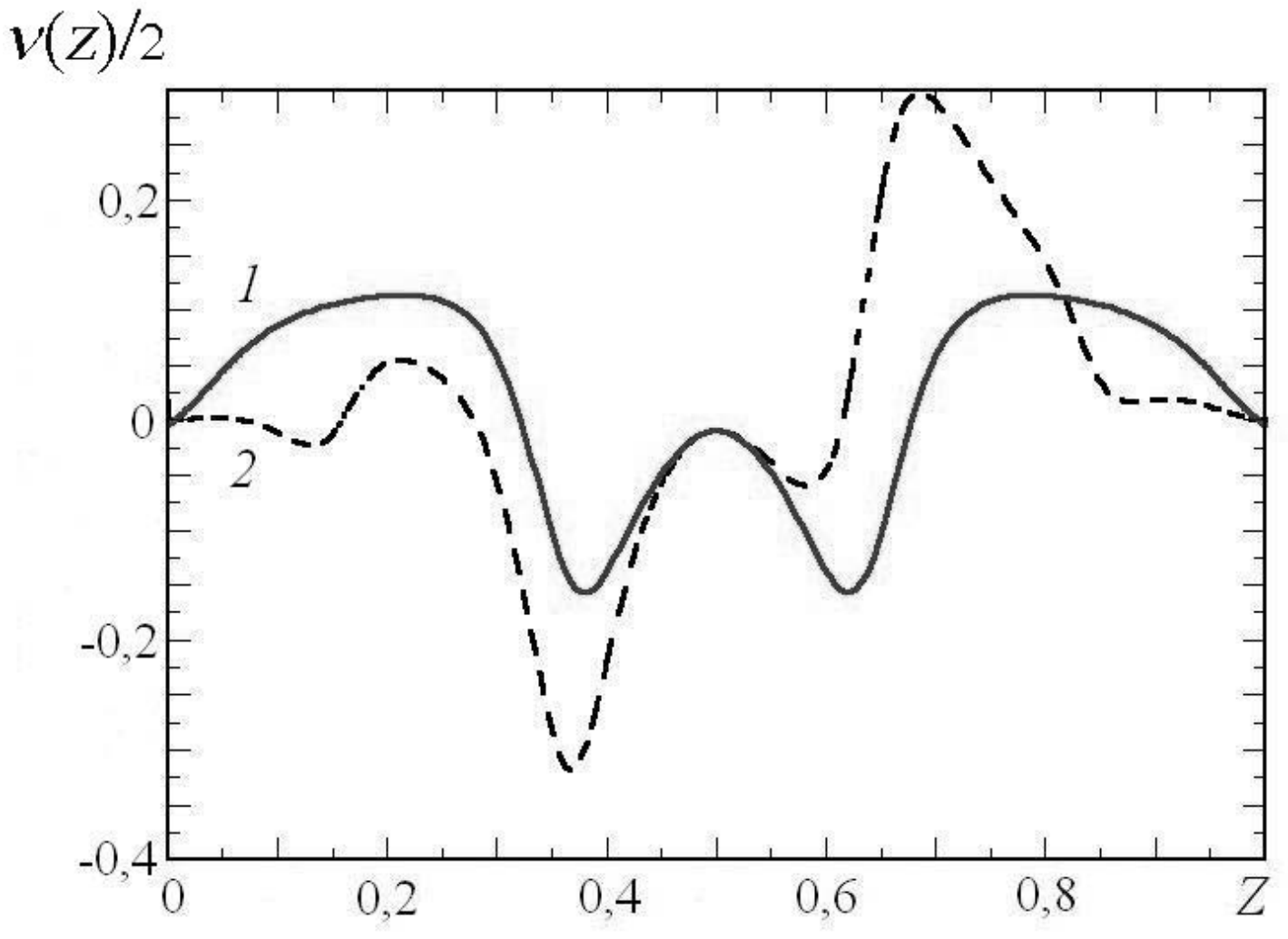

**Fig. 4.** Helicity for the purely *D* state of the deuteron versus the fraction *z* of the momentum of the system for the (*1* – solid line) nonrelativistic and (*2* – dashed line) relativistic cases.

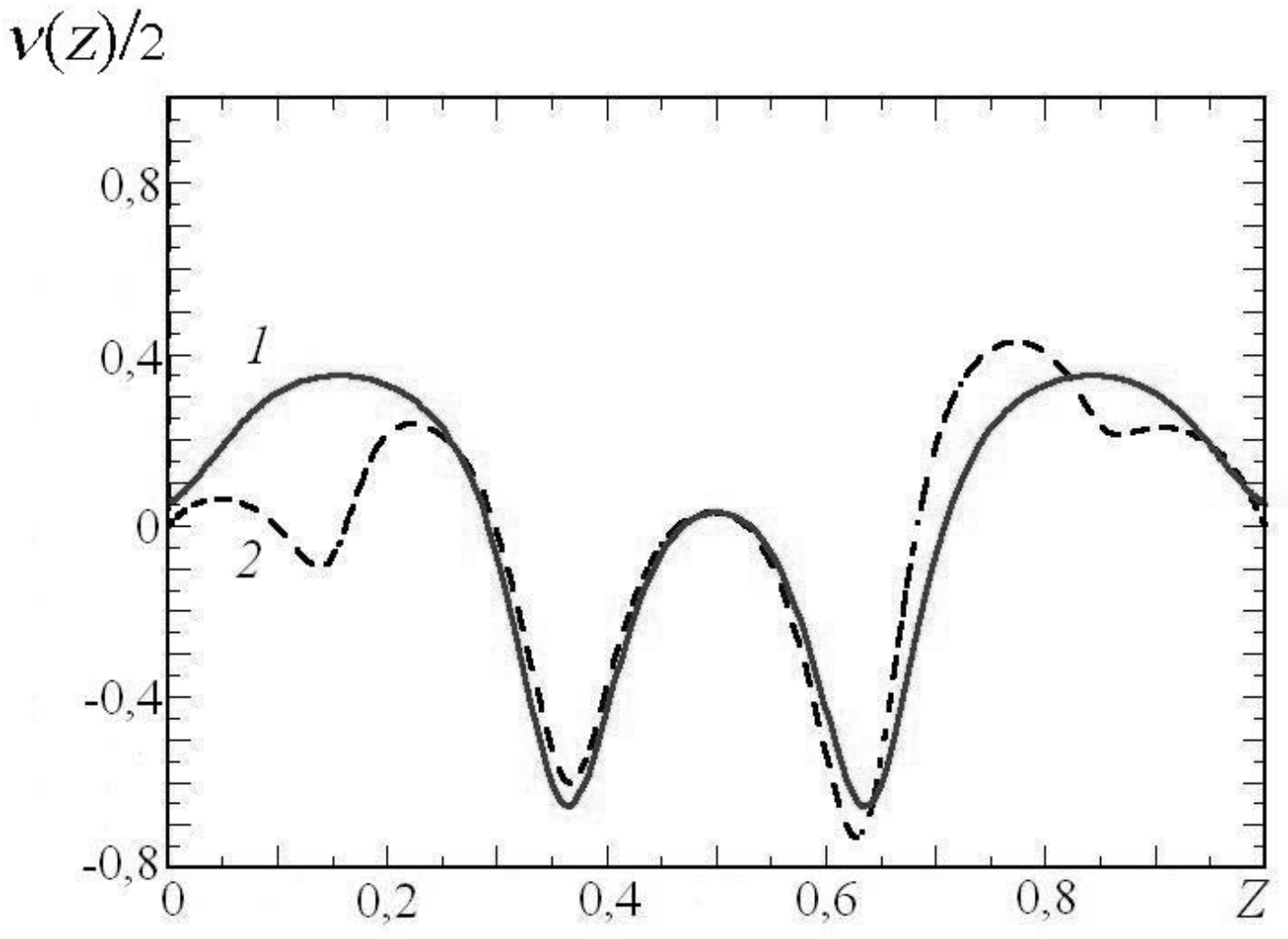

**Fig. 5.** Helicity for the purely *SD* (interference) state of the deuteron versus the fraction *z* of the momentum of the system for the (*1* – solid line) nonrelativistic and (*2* – dashed line) relativistic cases.

## VII. Relativistic Nuclear Corrections to the First Moment of the Spin-Dependent Structure Function of the Deuteron $\Gamma_1^D(Q^2)$

If the transverse momentum of a quark is negligible compared to its longitudinal momentum in the deep inelastic scattering of leptons on protons at high energies, the 4-momentum of the quark can be represented in the form $x_N p_\mu$, where $x_N = Q^2/2pq$ ($0 < x_N < 1$) is the Bjorken dimensionless scaling variable for the nucleon, $p_\mu$ is the 4-momentum of the nucleon, $q_\mu$ is the 4-momentum transferred by a virtual photon, and $Q^2 = -q^2$. Furthermore, if the 4-momentum of the quark is represented in the form $x_D P_\mu$, where $x_D = Q^2/2Pq$ and $P_\mu$ is the 4-momentum of the deuteron, then

$$p_\mu = (x_D / x_N) P_\mu,\ p_+ = (x_D / x_N) P_+ = zP_+,\ z = x_D / x_N . \tag{50}$$

As is known, the spin structure function of the nucleon $g_1^N(x_N, Q^2)$ is the difference between the probabilities that the quark in the longitudinally polarized nucleon has the momentum fraction $x_N$ and its spin is directed along and against the spin of the nucleon:

$$g_1(x_N) = \frac{1}{2}\sum_q e_q^2 \left[q^\uparrow(x_N) - q^\downarrow(x_N)\right], \tag{51}$$

where $e_q$ is the charge of the quark and $q^\uparrow(x_N)$ is the distribution of quarks with the spin projection +1/2 on the nucleon spin direction over the momentum fraction $x_N$.

The spin structure function of the nucleon $g_1^N(x_N, Q^2)$ can be represented in the form of the half-sum of the spin structure functions of the proton and neutron:

$$g_1^N(x_N, Q^2) = \frac{1}{2}\left(g_1^p(x_N, Q^2) + g_1^n(x_N, Q^2)\right). \tag{52}$$

In the first approximation, the spin structure function of the proton is obtained from experimentally observed quantities-longitudinal asymmetry $A_\parallel$, depolarization factor $D$ of the virtual photon, and unpolarized structure functions $F_2^p(x, Q^2)$ and $R(x, Q^2)$ [20]:

$$g_1^p = \frac{F_2^p\left(x,Q^2\right)}{2x\left(1+R\left(x,Q^2\right)\right)}\frac{A_{\parallel}}{D}. \quad (53)$$

The spin structure function of the neutron is extracted from the measured spin structure functions of the deuteron and proton [21].

The technique for the calculation of nuclear corrections to the structure function of the deuteron is topical, because the deuteron data are used to determine the spin structure function of the neutron, and was discussed in numerous works (see, e.g., [22–24]).

The spin structure function of the deuteron $g_1^D\left(x_N, Q^2\right)$ in the Bjorken limit is expressed in terms of the average helicity of the deuteron $\nu(z)$ and the spin structure function of the nucleon $g_1^N\left(x_N, Q^2\right)$ as [19]:

$$g_1^D\left(x_D, Q^2\right) = \int_{x_D}^{1} \frac{dz}{z}\nu(z) g_1^N\left(\frac{x_D}{z}, Q^2\right). \quad (54)$$

The first moment of the spin structure function of the deuteron is

$$\Gamma_1^D\left(Q^2\right) = \int_0^1 g_1^D\left(x_D, Q^2\right)dx_D = \int_0^1 dx_D \int_{x_D}^{1} \frac{dz}{z}\nu(z) g_1^N\left(\frac{x_D}{z}, Q^2\right), \quad (55)$$

and first moment of the spin structure function of the nucleon

$$\Gamma_1^N\left(Q^2\right) = \int_0^1 g_1^N\left(x_N, Q^2\right)dx_N \quad (56)$$

characterizes the total contribution of quarks to the spin of the nucleon.

With the change of the variables $x_N = x_D / z$ and Eq. (47), expression (55) for the first moment of the spin structure function of the deuteron can be separated into the nonrelativistic part and relativistic correction as

$$\Gamma_1^D\left(Q^2\right) = \int_0^1 \nu(z)dz \int_0^1 g_1^N\left(x_N, Q^2\right)dx_N = \left\langle \nu_p \right\rangle \int_0^1 g_1^N\left(x_N, Q^2\right)dx_N, \quad (57)$$

$$\Gamma_1^D\left(Q^2\right) = \left\langle \nu_p \right\rangle \Gamma_1^N\left(Q^2\right) = \left(1 - \frac{3}{2}w_D\right)\Gamma_1^N\left(Q^2\right) + \Delta_{rel}\Gamma_1^N\left(Q^2\right). \quad (58)$$

The experimental values of the first moments of the spin structure functions of the proton $\Gamma_1^p$ and neutron $\Gamma_1^n$ at $Q^2 = 5\ \text{GeV}^2$ obtained by the E155 collaboration by analyzing all available data are as follows [25]:

$$\Gamma_1^p = 0{,}118 \pm 0{,}004\ \text{(stat.)} \pm 0{,}007\ \text{(sist.)}, \tag{59}$$

$$\Gamma_1^n = -0{,}058 \pm 0{,}005\ \text{(stat.)} \pm 0{,}008\ \text{(sist.)}. \tag{60}$$

The experimental value for $\Gamma_1^D$ is

$$\Gamma_1^D = 0{,}028 \pm 0{,}004\ \text{(stat.)} \pm 0{,}005\ \text{(sist.)}. \tag{61}$$

The Bjorken and Ellis–Jaffe sum rules are theoretical relations of the first moments for the proton and neutron to the fundamental weak coupling constants. The test of the Bjorken sum rule for $Q^2 = 5\ \text{GeV}^2$ with the experimental data [25] gives

$$\Gamma_1^p - \Gamma_1^n = 0{,}176 \pm 0{,}003\ \text{(stat.)} \pm 0{,}007\ \text{(sist.)}. \tag{62}$$

The first moment of the spin structure function of the neutron extracted from the measured first moments of the spin structure functions of the deuteron $\Gamma_1^D\big|_{\exp}$ and proton $\Gamma_1^p\big|_{\exp}$ including the relativistic correction is represented in the form

$$\Gamma_1^n\left(Q^2\right) = \frac{2\Gamma_1^D\left(Q^2\right)\big|_{\exp}}{\left\langle \nu_p \right\rangle} - \Gamma_1^p\left(Q^2\right)\big|_{\exp}. \tag{63}$$

At $Q^2 = 5\ \text{GeV}^2$, $\Gamma_1^n$ = –0.05621. The Bjorken sum at $Q^2 = 5\ \text{GeV}^2$ gives $\Gamma_1^p - \Gamma_1^n$ = 0.17421.

There are several different parameterizations for the spin structure function of the nucleon $g_1^N\left(x, Q^2\right)$. In this work, the parameterization of parton distributions GRSV2000 [26], DNS2005 [27], LSS2006 [28] is used.

Figure 6*a*) shows the results of the calculation of the structure function of the nucleon given by Eq. (52) at $Q^2 = 5\ \text{GeV}^2$ and this function multiplied by $(1-(3/2)w_D)$ at $Q^2 = 5\ \text{GeV}^2$ parameterized in terms of parton distributions GRSV2000. Figure 6*b*) shows the structure function of the deuteron $g_1^D$ calculated by Eq. (54) at $Q^2 = 5\ \text{GeV}^2$ [19]. Figure 6*c*) shows the results of the calculation of the structure function of the nucleon given by Eq. (52) at $Q^2 = 5\ \mathrm{GeV}^2$ and this function multiplied by $(1-(3/2)w_D)$ at $Q^2 = 5\ \mathrm{GeV}^2$ parameterized in terms of parton distributions DNS2005. Figure 6*d*) shows the structure function of the deuteron $g_1^D$ calculated by Eq. (54) at $Q^2 = 5\ \mathrm{GeV}^2$ [19]. Figure 6*e*) shows the results of the calculation of the structure function of the nucleon given by Eq. (52) at $Q^2 = 5\ \mathrm{GeV}^2$ and this function multiplied by $(1-(3/2)w_D)$ at $Q^2 = 5\ \mathrm{GeV}^2$ parameterized in terms of parton distributions LSS2006. Figure 6*f*) shows the structure function of the deuteron $g_1^D$ calculated by Eq. (54) at $Q^2 = 5\ \mathrm{GeV}^2$ [19]. The results were obtained with the Bonn wave function of the deuteron [16].

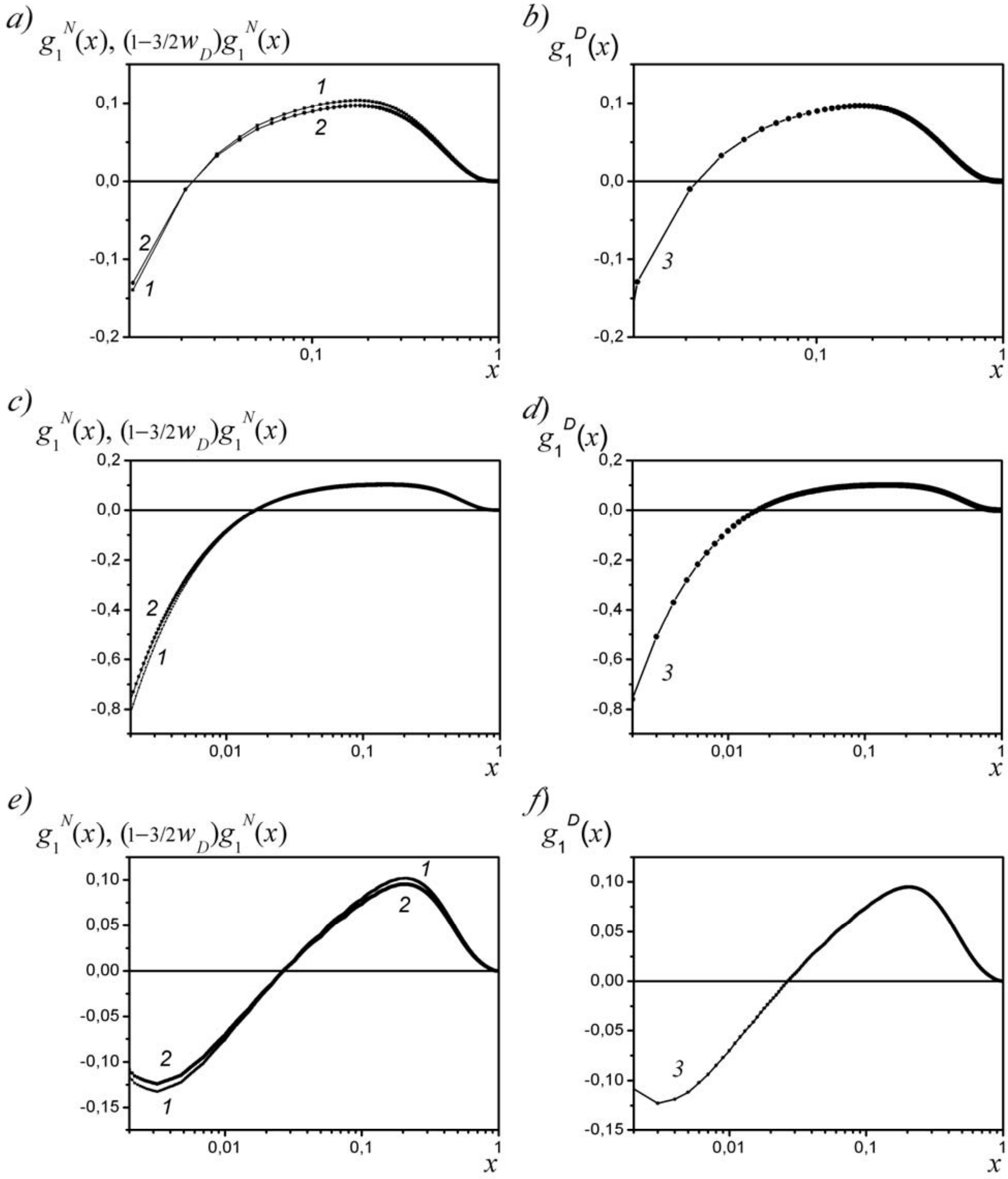

**Fig. 6.** Spin structure functions of (*a, c, e*) the nucleon (*1*) $g_1^N(x)$ and (*2*) $\left(1-(3/2)w_D\right)g_1^N(x)$ calculated by Eq. (52) and of (*b, d, f*) the deuteron $g_1^D$ calculated by Eq. (54) versus *x* for $Q^2 = 5\ \mathrm{GeV}^2$, parameterized in terms of parton distributions GRSV2000 (*a, b*), DNS2005 (*c*, *d*), LSS2006 (*e*, *f*).

According to Fig. 6, the spin structure function of the deuteron $g_1^D$ calculated by relativistic formula (54) as a function of $x$ slightly differs from the spin structure function $(1-(3/2)w_D)g_1^N$ parameterized in terms of parton distributions. It can be seen that the contribution of the relativistic correction to the spin structure function of the deuteron $g_1^D$ is small. This is due to the fact that the nonrelativistic wave function of the deuteron was used [16]. This contribution calculated with the wave function of the deuteron describing small nucleon–nucleon distances including contributions from the quark-gluon structure of the nucleon will expectedly not be so small. The total solution of the equation for the deuteron on the light cone has not yet been obtained. A number of widely used potentials contain components that cannot be treated in the field theory formalism. For this reason, relativistic effects have been estimated in the initial approximation with correspondence rules (33) and (34) and the modern realistic Bonn and Paris wave functions. A procedure for consistent calculation of relativistic nuclear corrections to the first moment of the spin-dependent structure function of deuteron is shown. In the relativistic case, the nucleon carrying a larger fraction $z$ of the momentum of the system makes a larger contribution to the distribution of the average helicity of the deuteron (Figures 2–5). This is an interesting result and deserves more discussion.

## ACKNOWLEDGMENTS

I am deeply grateful to N.N. Nikolaev (IKP FZJ and Landau Inst.) for valuable advices and discussions stimulating the appearance of this work, to S.I. Manaenkov (PNPI, Gatchina) for critical comments and useful discussions. The author wishes to thank Prof. J. Speth and Prof. Ulf-G. Meißner for the opportunity to be accepted as a collaborator at Theoretical group at Institute for Nuclear Physics at Forschungszentrum Jülich. I would also like to thank Igor Ivanov for many helpful discussions.

# Appendix A
## Dirac spinors

The Dirac spinor for the positive-frequency solution can be written as [29]:

$$u(p)=\begin{pmatrix}\sqrt{E+m}w\\ \sqrt{E-m}(\boldsymbol{\sigma}\cdot\mathbf{n})w\end{pmatrix}=\sqrt{E+m}\begin{pmatrix}w\\ \frac{\boldsymbol{\sigma}\cdot\mathbf{p}}{E+m}w\end{pmatrix}. \tag{A.1}$$

In the light-cone formalism to describe particles with spin 1/2 and 4-momentum $p=(p_+,p_-,\mathbf{p}_\perp)$, spinors are used [1, 7, 9]:

$$u(p,\lambda)=\frac{1}{\sqrt{\sqrt{2}p_+}}\left(\sqrt{2}p_++\beta m+\boldsymbol{\alpha}\cdot\mathbf{p}_\perp\right)\chi_\lambda, \tag{A.2}$$

where $\beta=\gamma_0=\begin{pmatrix}1&0\\0&-1\end{pmatrix}$, $\boldsymbol{\alpha}=\begin{pmatrix}0&\boldsymbol{\sigma}\\ \boldsymbol{\sigma}&0\end{pmatrix}$, $p_\pm=\frac{1}{\sqrt{2}}(p_0\pm p_3)$; the two $\chi_\lambda$-spinors for $\lambda=\pm1$ are

$$\chi_{\lambda=1}=\frac{1}{\sqrt{2}}\begin{pmatrix}1\\0\\1\\0\end{pmatrix},\quad \chi_{\lambda=-1}=\frac{1}{\sqrt{2}}\begin{pmatrix}0\\1\\0\\-1\end{pmatrix}. \tag{A.3}$$

Spinor $\mathrm{v}(p,\lambda)$ for antifermion

$$\mathrm{v}(p,\lambda)=\frac{1}{\sqrt{\sqrt{2}p_+}}\left(\sqrt{2}p_+-\beta m+\boldsymbol{\alpha}\cdot\mathbf{p}_\perp\right)\chi_{-\lambda}. \tag{A.4}$$

The explicit expressions for all spinors read:

$$u(p,\lambda=1)=N\begin{pmatrix}\sqrt{2}p_++m\\ p_x+ip_y\\ \sqrt{2}p_+-m\\ p_x+ip_y\end{pmatrix},\quad u(p,\lambda=-1)=N\begin{pmatrix}-p_x+ip_y\\ \sqrt{2}p_++m\\ p_x-ip_y\\ -\sqrt{2}p_++m\end{pmatrix},$$

$$\mathrm{v}(p,\lambda=1)=N\begin{pmatrix}-p_x+ip_y\\ \sqrt{2}p_+-m\\ p_x-ip_y\\ -\sqrt{2}p_+-m\end{pmatrix},\quad \mathrm{v}(p,\lambda=-1)=N\begin{pmatrix}\sqrt{2}p_+-m\\ p_x+ip_y\\ \sqrt{2}p_++m\\ p_x+ip_y\end{pmatrix}, \tag{A.5}$$

where $N=\frac{1}{\sqrt{2\sqrt{2}p_+}}$.

They are orthonormal and complete:

$\bar{u}(p,\lambda')u(p,\lambda)=-\bar{\text{v}}(p,\lambda)\text{v}(p,\lambda')=2m\delta_{\lambda\lambda'}$,

$\sum_{\lambda}u(p,\lambda)\bar{u}(p,\lambda)=\hat{p}+m$,

$$\sum_{\lambda}\text{v}(p,\lambda)\bar{\text{v}}(p,\lambda)=\hat{p}-m. \tag{A.6}$$

Since $(\boldsymbol{\sigma}\cdot\mathbf{n})X=\lambda X$, where **n** is the unit vector along the $z$-axis, spinor $u(p,\lambda)$ can be written as

$$u(p,\lambda=1)=N\begin{pmatrix}\sqrt{2}p_+ +m & (\boldsymbol{\sigma}\cdot\mathbf{p}_\perp)\\ (\boldsymbol{\sigma}\cdot\mathbf{p}_\perp) & \sqrt{2}p_+ -m\end{pmatrix}\begin{pmatrix}X\\ (\boldsymbol{\sigma}\cdot\mathbf{n})X\end{pmatrix}=N\begin{pmatrix}\left[\sqrt{2}p_+ +m+(\boldsymbol{\sigma}\cdot\mathbf{p}_\perp)(\boldsymbol{\sigma}\cdot\mathbf{n})\right]X\\ \left[(\boldsymbol{\sigma}\cdot\mathbf{p}_\perp)+\left(\sqrt{2}p_+ -m\right)(\boldsymbol{\sigma}\cdot\mathbf{n})\right]X\end{pmatrix}. \tag{A.7}$$

To establish a correspondence with the usual type of solutions of the Dirac equation (A.1)

$$u(p)=\sqrt{E+m}\begin{pmatrix}w\\ \frac{\boldsymbol{\sigma}\cdot\mathbf{p}}{E+m}w\end{pmatrix}=\sqrt{E+m}\begin{pmatrix}w\\ w'\end{pmatrix} \tag{A.8}$$

define the spinor $w=\left[\sqrt{2}p_+ +m+(\boldsymbol{\sigma}\cdot\mathbf{p}_\perp)(\boldsymbol{\sigma}\cdot\mathbf{n})\right]X$.

Then, we obtain

$$X=\frac{\sqrt{2}p_+ +m-(\boldsymbol{\sigma}\cdot\mathbf{p}_\perp)(\boldsymbol{\sigma}\cdot\mathbf{n})}{\left(\sqrt{2}p_+ +m\right)^2+\mathbf{p}_\perp^2}w=\frac{\sqrt{2}p_+ +m-(\boldsymbol{\sigma}\cdot\mathbf{p}_\perp)(\boldsymbol{\sigma}\cdot\mathbf{n})}{2\sqrt{2}p_+(E+m)}w. \tag{A.9}$$

Using this expression in the lower component spinor (A.7), after simple transformations we obtain:

$$w'=\frac{\left[(\boldsymbol{\sigma}\cdot\mathbf{p}_\perp)+\left(\sqrt{2}p_+ -m\right)(\boldsymbol{\sigma}\cdot\mathbf{n})\right]\left[\sqrt{2}p_+ +m-(\boldsymbol{\sigma}\cdot\mathbf{p}_\perp)(\boldsymbol{\sigma}\cdot\mathbf{n})\right]}{2\sqrt{2}p_+(E+m)}w=$$

$$=\frac{2\sqrt{2}p_+\left[(\boldsymbol{\sigma}\cdot\mathbf{p}_\perp)+(\boldsymbol{\sigma}\cdot\mathbf{n})p_z\right]}{2\sqrt{2}p_+(E+m)}w=\frac{\boldsymbol{\sigma}\cdot\mathbf{p}}{E+m}w. \tag{A.10}$$

Thus, the resulting spinor $X$ differs from the usual spinor $w$ only spin rotation, which is well-known transformation of the Wigner – Melosh [10, 14].

**Table I**

Matrix elements for the helicity spinors of Appendix A. Moreover every matrix element should be multiplied factor $\sqrt{k_+ p_+}$ .

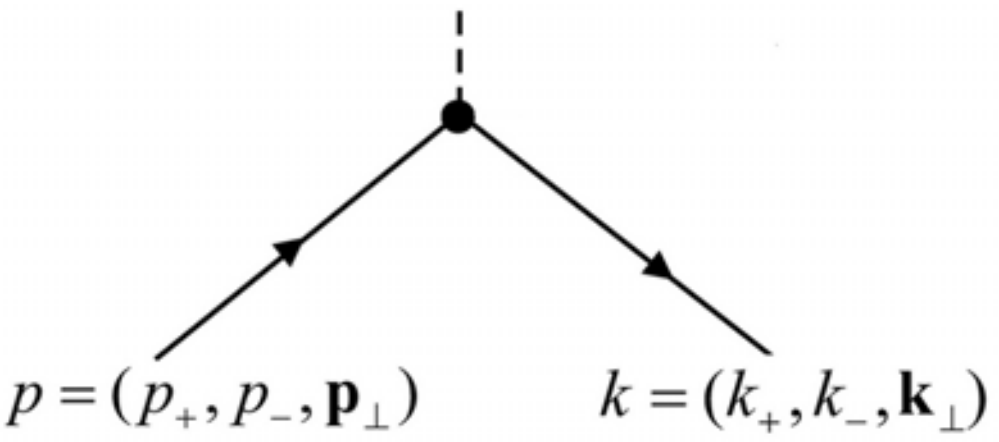

| $\Gamma$ | $\overline{u}(k,\lambda)\Gamma u(p,\lambda)$ | $\overline{u}(k,-\lambda)\Gamma u(p,\lambda)$ |
|---|---|---|
| $\gamma_+$ | $2$ | $0$ |
| $\gamma_-$ | $\frac{1}{k_+p_+}\left(m^2-k(-\lambda)p(\lambda)\right)$ | $\frac{m}{k_+p_+}\left(k(\lambda)-p(\lambda)\right)$ |
| $(\mathbf{a}\cdot\boldsymbol{\gamma})$ | $-\left(\frac{a(\lambda)k(-\lambda)}{k_+}+\frac{a(-\lambda)p(\lambda)}{p_+}\right)$ | $-ma(\lambda)\left(\frac{1}{k_+}-\frac{1}{p_+}\right)$ |
| $\gamma_x$ | $\lambda\left(\frac{k(-\lambda)}{k_+}-\frac{p(\lambda)}{p_+}\right)$ | $\lambda m\left(\frac{1}{k_+}-\frac{1}{p_+}\right)$ |
| $\gamma_y$ | $i\left(\frac{k(-\lambda)}{k_+}-\frac{p(\lambda)}{p_+}\right)$ | $im\left(\frac{1}{k_+}-\frac{1}{p_+}\right)$ |
| $I$ | $m\left(\frac{1}{k_+}+\frac{1}{p_+}\right)$ | $\frac{k(\lambda)}{k_+}-\frac{p(\lambda)}{p_+}$ |
| $\gamma_5=i\gamma_0\gamma_x\gamma_y\gamma_z$ | $\lambda m\left(\frac{1}{k_+}-\frac{1}{p_+}\right)$ | $\lambda\left(\frac{k(\lambda)}{k_+}-\frac{p(\lambda)}{p_+}\right)$ |
| $\gamma_+\gamma_5$ | $2\lambda$ | $0$ |
| $\gamma_-\gamma_5$ | $-\frac{\lambda}{k_+p_+}\left(m^2+k(-\lambda)p(\lambda)\right)$ | $-\frac{\lambda m}{k_+p_+}\left(k(\lambda)+p(\lambda)\right)$ |
| $(\mathbf{a}\cdot\boldsymbol{\gamma})\gamma_5$ | $-\lambda\left(\frac{a(\lambda)k(-\lambda)}{k_+}+\frac{a(-\lambda)p(\lambda)}{p_+}\right)$ | $-\lambda ma(\lambda)\left(\frac{1}{k_+}+\frac{1}{p_+}\right)$ |
| $\gamma_x\gamma_5$ | $\frac{k(-\lambda)}{k_+}-\frac{p(\lambda)}{p_+}$ | $m\left(\frac{1}{k_+}+\frac{1}{p_+}\right)$ |
| $\gamma_y\gamma_5$ | $i\lambda\left(\frac{k(-\lambda)}{k_+}+\frac{p(\lambda)}{p_+}\right)$ | $i\lambda m\left(\frac{1}{k_+}+\frac{1}{p_+}\right)$ |
| $\sigma_{+-}$ | $-im\left(\frac{1}{k_+}-\frac{1}{p_+}\right)$ | $-i\left(\frac{k(\lambda)}{k_+}+\frac{p(\lambda)}{p_+}\right)$ |
| $\sigma_{a+}$ | 0 | $2ia(\lambda)$ |
| $\sigma_{x+}$ | 0 | $-2i\lambda$ |
| $\sigma_{y+}$ | 0 | 2 |
| $\sigma_{a-}$ | $\frac{im}{k_+p_+}\left(-\frac{a(\lambda)k(-\lambda)}{k_+}+\frac{a(-\lambda)p(\lambda)}{p_+}\right)$ | $\frac{i}{k_+p_+}\left(-m^2a(\lambda)+a(-\lambda)k(\lambda)p(\lambda)\right)$ |
| $\sigma_{ab}$ | $\lambda m[\mathbf{a},\mathbf{b}]\left(\frac{1}{k_+}+\frac{1}{p_+}\right)$ | $\lambda[\mathbf{a},\mathbf{b}]\left(\frac{k(\lambda)}{k_+}+\frac{p(\lambda)}{p_+}\right)$ |
| $\sigma_{xy}$ | $\lambda m\left(\frac{1}{k_+}+\frac{1}{p_+}\right)$ | $\lambda\left(\frac{k(\lambda)}{k_+}+\frac{p(\lambda)}{p_+}\right)$ |
| $(\mathbf{a}\cdot\boldsymbol{\gamma})\gamma_+\gamma_5$ | 0 | $2\lambda a(\lambda)$ |
| $\gamma_x\gamma_+\gamma_5$ | 0 | $-2$ |
| $\gamma_y\gamma_+\gamma_5$ | 0 | $-2i\lambda$ |
| $\sigma_{a+}\gamma_5$ | 0 | $2i\lambda a(\lambda)$ |
| $\sigma_{x+}\gamma_5$ | 0 | $-2i$ |
| $\sigma_{y+}\gamma_5$ | 0 | $2\lambda$ |

For $\lambda=\pm 1$ we defined

$$p(\lambda)\equiv-\lambda p_x-ip_y\,,\;k(-\lambda)\equiv\lambda k_x-ik_y\,.$$

Some useful relations:

$$k(-\lambda)p(\lambda)=-\mathbf{p}_\perp\cdot\mathbf{k}_\perp-i\lambda[\mathbf{k}_\perp,\mathbf{p}_\perp]_z\,,$$

$$[\mathbf{k}_\perp,\mathbf{p}_\perp]_z=k_xp_y-k_yp_x\,,\;\mathbf{p}_\perp=(p_x,\,p_y)\,,\;\mathbf{k}_\perp=(k_x,\,k_y)\,,$$

$$(\mathbf{a}\cdot\boldsymbol{\gamma})=a_x\gamma_x+a_y\gamma_y\,,$$

$$\sigma_{ab}=\frac{i}{2}\left[(\mathbf{a}\cdot\boldsymbol{\gamma})(\mathbf{b}\cdot\boldsymbol{\gamma})-(\mathbf{b}\cdot\boldsymbol{\gamma})(\mathbf{a}\cdot\boldsymbol{\gamma})\right],\;\;\sigma_{a+}=\frac{i}{2}\left[(\mathbf{a}\cdot\boldsymbol{\gamma})\gamma_+-\gamma_+(\mathbf{a}\cdot\boldsymbol{\gamma})\right]=a_x\sigma_{x+}+a_y\sigma_{y+}\,.$$

**Table II**

Matrix elements for the helicity spinors of Appendix A. Moreover every matrix element should be multiplied factor $\sqrt{p_{1+}p_{3+}}$ .

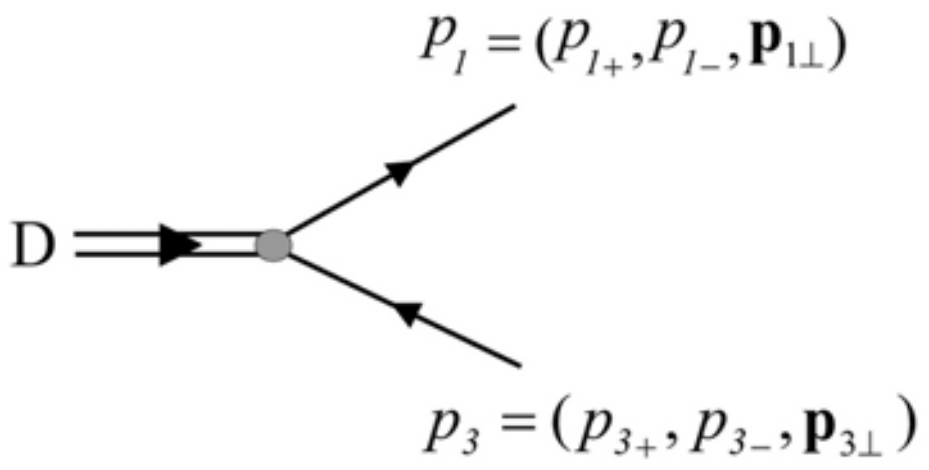

| $\Gamma$ | $\bar{u}(p_1,\lambda)\Gamma \mathrm{v}(p_3,\lambda)$ | $\bar{u}(p_1,\lambda)\Gamma \mathrm{v}(p_3,-\lambda)$ |
|---|---|---|
| $\gamma_+$ | 0 | 2 |
| $\gamma_-$ | $-\frac{m}{p_{1+}p_{3+}}\left(p_1(-\lambda)+p_3(-\lambda)\right)$ | $-\frac{1}{p_{1+}p_{3+}}\left(m^2+p_1(-\lambda)p_3(\lambda)\right)$ |
| $(\mathbf{a}\cdot\boldsymbol{\gamma})$ | $-ma(-\lambda)\left(\frac{1}{p_{1+}}+\frac{1}{p_{3+}}\right)$ | $-\left(\frac{a(\lambda)p_1(-\lambda)}{p_{1+}}+\frac{a(-\lambda)p_3(\lambda)}{p_{3+}}\right)$ |
| $I$ | $\frac{p_1(-\lambda)}{p_{1+}}-\frac{p_3(-\lambda)}{p_{3+}}$ | $m\left(\frac{1}{p_{1+}}-\frac{1}{p_{3+}}\right)$ |

**Table III**

Matrix elements for the helicity spinors of Appendix A. Moreover every matrix element should be multiplied factor $\sqrt{p_{3+}p_{2+}}$ .

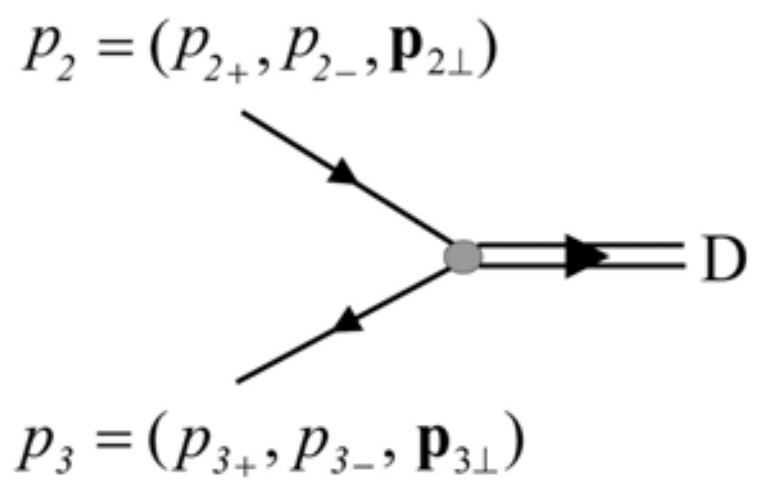

| $\Gamma$ | $\bar{v}(p_3,\lambda)\Gamma u(p_2,\lambda)$ | $\bar{v}(p_3,-\lambda)\Gamma u(p_2,\lambda)$ |
|---|---|---|
| $\gamma_+$ | 0 | 2 |
| $\gamma_-$ | $\frac{m}{p_{2+}p_{3+}}\left(p_3(\lambda)+p_2(\lambda)\right)$ | $-\frac{1}{p_{2+}p_{3+}}\left(m^2+p_3(-\lambda)p_2(\lambda)\right)$ |
| $(\mathbf{a}\cdot\boldsymbol{\gamma})$ | $ma(\lambda)\left(\frac{1}{p_{3+}}+\frac{1}{p_{2+}}\right)$ | $-\left(\frac{a(\lambda)p_3(-\lambda)}{p_{3+}}+\frac{a(-\lambda)p_2(\lambda)}{p_{2+}}\right)$ |
| $I$ | $\frac{p_3(\lambda)}{p_{3+}}-\frac{p_2(\lambda)}{p_{2+}}$ | $m\left(-\frac{1}{p_{3+}}+\frac{1}{p_{2+}}\right)$ |

# Appendix B

## Polarization vectors in light-cone gauge

Let us consider the Breit frame in which the deuteron moves along the *x*-axis. Futher we choose 4-momentum of the deuteron (two-nucleon Fock states) in Minkowski space-time with one temporal and three spatial components in the form:

$$P=(P_0,P_1,P_2,P_3)=(P_0,P_x,0,0), \tag{B.1}$$

or, in the light-cone formalism, with the plus, minus and transverse components:

$$P=(P_+,P_-,P_1,P_2)=\left(P_+,\frac{M_\perp^2}{2P_+},P_x,0\right), \tag{B.2}$$

where

$$M_\perp^2=M^2+P_x^2. \tag{B.3}$$

4-vectors of polarization of the deuteron (two-nucleon Fock states) in Minkowski space-time with one temporal and three spatial components are given by

$$V^{/(y)}=(0,0,1,0), \tag{B.4}$$

$$V^{/(z)}=(0,0,0,1), \tag{B.5}$$

$$V^{/(x)}=\frac{1}{M}(P_x,M_\perp,0,0), \tag{B.6}$$

or, in the representation, where the 4-vector has light-cone components:

$$V^{/(y)}=(0,0,0,1), \tag{B.7}$$

$$V^{/(z)}=\left(\frac{1}{\sqrt{2}},-\frac{1}{\sqrt{2}},0,0\right), \tag{B.8}$$

$$V^{/(x)}=\frac{1}{M}\left(\frac{P_x}{\sqrt{2}},\frac{P_x}{\sqrt{2}},M_\perp,0\right). \tag{B.9}$$

Consider the frame in which $P_z\neq 0$. Let us fulfill the boost along the *z*-axis with $\gamma$-factor $\sqrt{2}P_+/M_\perp$.

Then,

$$P=\left(P_+,\frac{M_\perp^2}{2P_+},\mathbf{P}_\perp\right), \tag{B.10}$$

$$V^{//(x)} = \frac{1}{M}\left( P_x \frac{P_+}{M_\perp}, P_x \frac{M_\perp}{2P_+}, M_\perp, 0 \right), \quad \text{(B.11)}$$

$$V^{//(y)} = (0, 0, 0, 1), \quad \text{(B.12)}$$

$$V^{//(z)} = \left( \frac{P_+}{M_\perp}, -\frac{M_\perp}{2P_+}, 0, 0 \right). \quad \text{(B.13)}$$

After the boost along the $z$-axis 4-vectors $V^{//(x)}$ and $V^{//(z)}$ acquire a temporary components and non-plus-zero components. It is convenient to choose a basis in which $V^{//(x)}$ does not have a "plus" component (light-cone gauge). For this let us perform the linear transformation in the form:

$$V^{(x)} = V^{//(x)} \cos\alpha + V^{//(z)} \sin\alpha,$$

$$V^{(z)} = -V^{//(x)} \sin\alpha + V^{//(z)} \cos\alpha, \quad \text{(B.14)}$$

where $\cos\alpha = M / M_\perp$, $\sin\alpha = -P_x / M_\perp$.

Then,

$$V^{(x)} = \left( 0, \frac{P_x}{P_+}, 1, 0 \right), \quad \text{(B.15)}$$

$$V^{(z)} = \frac{1}{M}\left( P_+, \frac{-M^2 + \mathbf{P}_\perp^2}{2P_+}, \mathbf{P}_\perp \right). \quad \text{(B.16)}$$

In the helicity representation

$$V^{(\rho=\pm1)} = -\frac{1}{\sqrt{2}}\left( \pm V^{(x)} + iV^{(y)} \right). \quad \text{(B.17)}$$

Then, 4-vectors of polarization of the deuteron (two-nucleon Fock states) in light-cone gauge are as follows:

$$V^{(\rho=\pm1)} = \left( 0, \frac{\left(\mathbf{P}_\perp \cdot \mathbf{e}^{(\rho=\pm1)}\right)}{P_+}, \mathbf{e}^{(\rho=\pm1)} \right), \quad \text{(B.18)}$$

$$V^{(\rho=0)} = V^{(z)} = \frac{1}{M}\left( P_+, \frac{-M^2 + \mathbf{P}_\perp^2}{2P_+}, \mathbf{P}_\perp \right). \quad \text{(B.19)}$$